\documentclass[
aps,
prd,
twocolumn,
superscriptaddress,
nofootinbib,
floatfix
]{revtex4-2}

\usepackage[T1]{fontenc}
\usepackage[utf8]{inputenc}
\usepackage{amsmath,amssymb,amsfonts}
\usepackage{bm}
\usepackage{physics}
\usepackage{placeins}
\usepackage{flafter}
\usepackage{float}
\usepackage{graphicx}
\usepackage{xcolor}
\usepackage{hyperref}

\hypersetup{
	colorlinks=true,
	linkcolor=blue,
	citecolor=blue,
	urlcolor=blue,
	pdfauthor={Hindi. Zouhair},
	pdftitle={Probing Lepton-Flavor Universality Violation in Semileptonic Heavy-Baryon Decays}
}

\newcommand{\Lambdab}{\Lambda_b}
\newcommand{\Lambdac}{\Lambda_c}

\newcommand{\Rlam}{R_{\Lambdac}}
 
\begin{document}
	
	\title{Lifting Effective-Field-Theory Degeneracies in Semileptonic Heavy-Baryon Decays}
	\author{Hindi~Zouhair}
	\email{hindizouhair@gmail.com}
	\affiliation{Laboratory of High Energy Physics (LHEP-MS), Mohammed V University, Rabat, Morocco}
	
	\date{\today}
	
	\begin{abstract}
		Semileptonic heavy-baryon decays provide a sensitive probe of the
		helicity structure underlying possible lepton-flavor universality
		violation in $b\to c\,\tau\bar\nu_\tau$ transitions.
		We perform an effective-field-theory analysis of
		$\Lambda_b\to\Lambda_c\tau\bar\nu_\tau$ and related baryonic modes
		using lattice-QCD helicity form factors with full covariance
		propagation. Propagating meson-compatible EFT solutions into the baryonic observable
		space$(R_{\Lambda_c},\,P_\tau,\,A_{\rm FB})$, we show that tau polarization provides the leading source of sensitivity for
		lifting EFT degeneracies that remain unresolved in current measurements
		of $R(D)$ and $R(D^\ast)$.
		Vector-like and tensor-like solutions remain clustered near the
		Standard-Model prediction, whereas scalar-containing directions produce
		large polarization displacements and characteristic low-$q^2$
		deformations of the normalized differential spectrum.
		A covariance-aware analysis demonstrates that baryonic polarization and
		differential observables provide complementary and independent
		information on the Lorentz structure of semileptonic new physics.
		$P_\tau(\Lambda_b\to\Lambda_c\tau\bar\nu_\tau)$
		and the low-$q^2$ spectrum as particularly powerful probes for future
		tests of semitauonic flavor anomalies at LHCb and future flavor
		facilities.
	\end{abstract}
	\maketitle
	
\section{Introduction}
\label{sec:intro}
Hints of lepton-flavor universality (LFU) violation in semileptonic
$b \to c\,\ell\,\bar\nu_\ell$ transitions, as indicated by measurements of
$R_D$ and $R_{D^*}$, constitute one of the most persistent tensions
with the Standard Model in flavor physics.
Current world averages show deviations at the level of a few standard
deviations, motivating extensive theoretical and experimental scrutiny
of these observables~\cite{BaBar2012,Belle2015,LHCb2015RDst,Belle2019,LHCb2023,HFLAV2024,BernlochnerRMP2022}.
While recent updates have reduced the overall significance, a wide
class of new-physics scenarios remains viable motivating increasingly precise EFT-global analyses of the surviving parameter space~\cite{Bordone2017}.

Most theoretical studies to date have focused on mesonic decays,
in particular $B\to D^{(*)}\ell\nu$ transitions, where global analyses
within an effective-field-theory framework reveal approximate
 degeneracies among vector-like and scalar operator directions when confronted with existing data~\cite{Sakaki2013,Freytsis2015,Alonso2017,Celis2017,Feruglio2017,Iguro2024}.
These degeneracies limit the discriminating power of mesonic observables
alone and motivate the exploration of additional, independent probes. This feature has emerged as a central obstacle in modern semitauonic EFT analyses~\cite{Murgui2019,Jung2019}.

Heavy-baryon semileptonic decays provide such a probe.
Processes such as $\Lambda_b \to \Lambda_c \ell \bar\nu_\ell$ and
$\Xi_b \to \Xi_c \ell \bar\nu_\ell$ involve spin-$1/2$ initial and final states,
leading to a rich helicity structure and direct access to polarization
and angular observables~\cite{Korner1992,Gutsche2015,Datta2017,Bernlochner2018Lambda}.
These features enable sensitivity to interference patterns and chiral
structures that are suppressed or inaccessible in mesonic decays.
The helicity organization of heavy-baryon amplitudes is particularly transparent in HQET-inspired formulations~\cite{Mannel1991}.

The central aim of this work is to assess the extent to which baryonic
observables can discriminate among new-physics scenarios compatible
with current measurements of $R_D$ and $R_{D^*}$.
In particular, polarization observables in
$\Lambda_b\to\Lambda_c\tau\bar\nu_\tau$ provide enhanced sensitivity to
non–Standard-Model helicity structures, complementing and extending
mesonic analyses. A key emphasis of our study is the propagation of hadronic uncertainties,
using lattice-QCD form factors~\cite{Detmold2015,Datta2017,Bernlochner2018Lambda},
to assess the robustness of these conclusions.

We perform a comprehensive analysis within a dimension-six effective
field theory, considering lepton-flavor-universality ratios,
polarization observables, and angular asymmetries in heavy-baryon decays.
Benchmark scenarios consistent with current global fits are explored,
and correlations with mesonic observables are identified.
Finally, we discuss the experimental sensitivity required at LHCb Run~3
and the LHCb Upgrade to exploit these observables~\cite{LHCbLambdaMu2017,LHCbLambdaTau2022,LHCbUpgrade2018}.
\paragraph{Mesonic acceptance and the origin of EFT degeneracies.}
To make the notion of ``meson-only degeneracies'' precise, we define a
\emph{mesonic-accepted} region in Wilson-coefficient space by requiring agreement
with the measured mesonic LFU ratios.
Concretely, we impose a $\chi^2$ criterion on the input set
$\{R(D),\,R(D^\ast)\}$ (and, where stated, additional information from
polarization and angular observables such as $P_\tau(D^\ast)$ and $F_L(D^\ast)$ ~\cite{Belle2019,HFLAV2024,Iguro2024}.
For an observable vector $\vec O$ with experimental covariance matrix $C_{\rm exp}$,
we define
\begin{equation}
\chi^2_{\rm meson} =
(\vec O^{\,\rm th}-\vec O^{\,\rm exp})^T\,C_{\rm exp}^{-1}\,(\vec O^{\,\rm th}-\vec O^{\,\rm exp}),
\label{eq:chi2_meson}
\end{equation}
and accept EFT points satisfying
\begin{equation}
\chi^2_{\rm meson} \le \chi^2_{n,\,0.9973},
\label{eq:chi2_cut_3sigma}
\end{equation}
where $n$ is the number of mesonic inputs and $\chi^2_{n,\,0.9973}$ corresponds to
a two-sided $3\sigma$ confidence region.
The resulting accepted set typically contains disconnected regions and approximate
flat directions (notably scalar-dominated correlated directions) that reproduce mesonic rates but differ in helicity structure.
The central purpose of this work is to show that baryonic observables break these
degeneracies in a robust manner once lattice-QCD uncertainties are propagated.
We include vector, scalar, tensor, and correlated scalar--tensor EFT
directions and propagate the meson-accepted parameter points to baryonic
observables using lattice-QCD form-factor inputs.

\paragraph{Core claim and strategy.}
The central claim of this work is that semileptonic heavy-baryon decays
provide observables that \emph{lift the approximate EFT degeneracies}
present in mesonic analyses of $b \to c\,\ell\,\bar\nu_\ell$.
Concretely, we show that the baryonic LFU ratio $R_{\Lambdac}$ together with
polarization and angular observables (notably $P_\tau$ and $A_{\rm FB}$)
exhibit \emph{distinct operator fingerprints} for vector-like and
correlated scalar interactions that remain compatible with current
$R_D$ and $R_{D^*}$ data.
This discrimination persists after consistently propagating
lattice-QCD form-factor uncertainties, demonstrating that baryonic
observables provide complementary sensitivity to the Lorentz structure
of semileptonic new physics beyond that accessible through mesonic
channels alone.

The remainder of this paper is organized as follows.
In Sec.~\ref{sec:eft} we introduce the effective-field-theory framework.
Section~\ref{sec:ff} summarizes the hadronic matrix elements and
lattice-QCD inputs, while the physical observables are defined in
Sec.~\ref{sec:obs}.
The numerical framework and uncertainty propagation are presented in
Sec.~\ref{sec:num}.
Results for integrated and differential baryonic observables are
discussed in Sec.~\ref{sec:results}, followed by the phenomenological
interpretation and experimental implications in
Sec.~\ref{sec:disc}.
We conclude in Sec.~\ref{sec:concl}.
\section{Effective field theory framework}
\label{sec:eft}
At energies well below the electroweak scale, semileptonic
$b \to c\,\ell\,\bar\nu_\ell$ transitions are described by an effective
Hamiltonian constructed from dimension-six operators
~\cite{Sakaki2013,Freytsis2015,Alonso2017,Celis2017}.
Assuming left-handed neutrinos and neglecting operators suppressed by
additional powers of light fermion masses, we write
\begin{equation}
\mathcal H_{\rm eff}
= \frac{4 G_F}{\sqrt{2}}\,V_{cb}\!
\left[
\begin{aligned}
&(1+g_{V_L})\,\mathcal O_{V_L}
+ g_{V_R}\,\mathcal O_{V_R}
+ g_{S_L}\,\mathcal O_{S_L} \\
&\quad
+ g_{S_R}\,\mathcal O_{S_R}
+ g_T\,\mathcal O_T
\end{aligned}
\right].
\label{eq:Heff_short}
\end{equation}
The coefficients in Eq.~\eqref{eq:Heff_short}
can be viewed as low-energy EFT remnants of
dimension-six SMEFT operators after electroweak matching and
renormalization-group evolution
~\cite{Jenkins2013,Alonso2014}.

The dimension-six operators appearing in
Eq.~\eqref{eq:Heff_short} are defined as
\begin{align}
\mathcal O_{V_L}
&=
(\bar c \gamma^\mu P_L b)
(\bar\ell \gamma_\mu P_L \nu),
\nonumber\\
\mathcal O_{V_R}
&=
(\bar c \gamma^\mu P_R b)
(\bar\ell \gamma_\mu P_L \nu),
\nonumber\\
\mathcal O_{S_L}
&=
(\bar c P_L b)
(\bar\ell P_L \nu),
\nonumber\\
\mathcal O_{S_R}
&=
(\bar c P_R b)
(\bar\ell P_L \nu),
\nonumber\\
\mathcal O_T
&=
(\bar c \sigma^{\mu\nu} P_L b)
(\bar\ell \sigma_{\mu\nu} P_L \nu).
\label{eq:Ops}
\end{align}

The tensor interaction is particularly sensitive to
renormalization-group evolution, since QCD running can significantly
modify the low-energy normalization of tensor Wilson
coefficients~\cite{GonzalezAlonso2017}.
Here
$P_{L,R}=(1\mp\gamma_5)/2$,
while the Wilson coefficients $g_i$ encode possible
non-Standard-Model contributions.
In the Standard Model,
$g_i=0$.
This operator basis coincides with that commonly employed in analyses
of $R_D$ and $R_{D^*}$ and facilitates a direct comparison between
mesonic and baryonic observables
~\cite{Sakaki2013,Alonso2017,Celis2017,Feruglio2017,Iguro2024}.
While mesonic decay rates are primarily sensitive to specific linear
combinations of the coefficients in
Eq.~\eqref{eq:Heff_short},
baryonic processes exhibit a richer helicity structure.
As a result, polarization and angular observables in
$\Lambda_b \to \Lambda_c \ell \bar\nu_\ell$
and
$\Xi_b \to \Xi_c \ell \bar\nu_\ell$
provide independent sensitivity to scalar and tensor interactions
through characteristic angular and polarization interference
patterns~\cite{Boer2019Lambda},
allowing degeneracies present in mesonic fits to be lifted.
\section{Hadronic matrix elements and form factors}
\label{sec:ff}
The hadronic dynamics of semileptonic heavy-baryon decays is encoded in
matrix elements of quark currents between spin-$1/2$ baryon states.
In this work we employ a helicity-based form-factor decomposition,
which is particularly well suited for incorporating lattice-QCD
determinations and for constructing polarization and angular observables.
Our conventions follow those adopted in state-of-the-art lattice and
phenomenological analyses of $\Lambda_b\to\Lambda_c\tau\bar\nu_\tau$
decays~\cite{Detmold2015,Datta2017,Bernlochner2018Lambda}.

The physical kinematic range of the decay is
\begin{equation}
m_\ell^2 \le q^2 \le (m_{\Lambdab}-m_{\Lambdac})^2 ,
\end{equation}
where the upper endpoint corresponds to the zero-recoil configuration
of the final-state baryon.
We define the momentum combinations
$q^\mu = p^\mu - p'^\mu$ and $P^\mu = p^\mu + p'^\mu$ throughout.
\subsection{Vector and axial-vector currents}
\label{subsec:ff_decomp}

The vector-current matrix element is written as
\begin{equation}
\langle \Lambdac | \bar c \gamma^\mu b | \Lambdab \rangle
=
\bar u_{\Lambdac}\, \Gamma_V^\mu(q^2)\, u_{\Lambdab} ,
\end{equation}
with
\begin{equation}
\begin{aligned}
\Gamma_V^\mu(q^2) &=
f_0(q^2)\,
\frac{m_{\Lambdab}-m_{\Lambdac}}{q^2}\, q^\mu
+ f_+(q^2)\,
\frac{m_{\Lambdab}+m_{\Lambdac}}{s_+}\, P^\mu  \\
&\quad
+ f_\perp(q^2)
\left(
\gamma^\mu - \frac{2 m_{\Lambdac}}{s_+} p^\mu
\right),
\end{aligned}
\end{equation}
where $s_+ = (m_{\Lambdab}+m_{\Lambdac})^2 - q^2$.

Similarly, the axial-vector current matrix element is given by
\begin{equation}
\langle \Lambdac | \bar c \gamma^\mu \gamma_5 b | \Lambdab \rangle
=
\bar u_{\Lambdac}\, \Gamma_A^\mu(q^2)\, \gamma_5\, u_{\Lambdab} ,
\end{equation}
with
\begin{equation}
\begin{aligned}
\Gamma_A^\mu(q^2) &=
g_0(q^2)\,
\frac{m_{\Lambdab}+m_{\Lambdac}}{q^2}\, q^\mu
+ g_+(q^2)\,
\frac{m_{\Lambdab}-m_{\Lambdac}}{s_-}\, P^\mu  \\
&\quad
+ g_\perp(q^2)
\left(
\gamma^\mu + \frac{2 m_{\Lambdac}}{s_-} p^\mu
\right),
\end{aligned}
\end{equation}
where $s_- = (m_{\Lambdab}-m_{\Lambdac})^2 - q^2$.
The form factors $f_i(q^2)$ and $g_i(q^2)$ are real functions of $q^2$
determined nonperturbatively from lattice-QCD calculations.
\subsection{Scalar and tensor currents}
\label{subsec:scalar_tensor}

The scalar and pseudoscalar matrix elements are related to the
vector and axial-vector currents through the equations of motion,
up to corrections proportional to light-quark masses.

The tensor-current matrix element is parameterized as
\begin{equation}
\langle \Lambdac | \bar c \sigma^{\mu\nu} b | \Lambdab \rangle
=
\bar u_{\Lambdac}\, \Gamma_T^{\mu\nu}(q^2)\, u_{\Lambdab} ,
\end{equation}
with
\begin{equation}
\Gamma_T^{\mu\nu}(q^2) =
h_+(q^2)\, \sigma^{\mu\nu}
+ h_\perp(q^2)\,
\frac{p^\mu q^\nu - p^\nu q^\mu}{m_{\Lambdab}} .
\end{equation}
The tensor form factors provide direct sensitivity to tensor
interactions beyond the Standard Model~\cite{Datta2017,Bernlochner2018Lambda}.
\subsection{Helicity amplitudes}
\label{subsec:hel}
Physical observables are expressed in terms of helicity amplitudes,
defined by contracting the hadronic matrix elements with polarization
vectors of the off-shell $W^\ast$ boson.
For the vector and axial-vector currents we define
\begin{align}
H^{V}_{\lambda_{\Lambdac},\lambda}(q^2)
&\equiv
\varepsilon_\mu^*(\lambda)\,
\langle \Lambdac(\lambda_{\Lambdac}) |
\bar c \gamma^\mu b |
\Lambdab \rangle ,
\label{eq:HV_def}
\\
H^{A}_{\lambda_{\Lambdac},\lambda}(q^2)
&\equiv
\varepsilon_\mu^*(\lambda)\,
\langle \Lambdac(\lambda_{\Lambdac}) |
\bar c \gamma^\mu \gamma_5 b |
\Lambdab \rangle ,
\label{eq:HA_def}
\end{align}
where $\lambda=\pm,0,t$ labels the polarization of the virtual $W^\ast$
($t$ denotes the timelike polarization).
For tensor interactions we introduce the helicity amplitudes
\begin{equation}
H^{T}_{\lambda_{\Lambdac},\lambda\lambda'}(q^2)
\equiv
\varepsilon_\mu^*(\lambda)\,
\varepsilon_\nu^*(\lambda')\,
\langle \Lambdac(\lambda_{\Lambdac}) |
\bar c \sigma^{\mu\nu} b |
\Lambdab \rangle .
\label{eq:HT_def}
\end{equation}
Explicit expressions of the helicity amplitudes in terms of the form
factors introduced in Sec.~\ref{sec:ff} are collected in Appendix~A.
All decay rates, polarization observables, and angular asymmetries
considered in this work are constructed from these amplitudes.
\subsection{Lattice-QCD inputs and uncertainty treatment}
\label{subsec:lattice_inputs}
For the numerical analysis we employ the lattice-QCD determination of
the $\Lambdab\to\Lambdac$ form factors and their full covariance matrix
from Ref.~\cite{Detmold2015}.
The published covariance information contains nontrivial correlations
among different form factors and kinematic points.
To consistently propagate hadronic uncertainties, the form-factor
parameters are sampled from the corresponding multivariate Gaussian
distribution defined by the lattice covariance matrix.
The correlated form-factor uncertainties are then propagated through
the full helicity-amplitude framework to all derived observables,
including $R_{\Lambda_c}$, $P_\tau$, $A_{\rm FB}$, and the differential
spectral observables discussed below.
This procedure preserves the complete correlation structure of the
lattice-QCD inputs and provides a statistically consistent estimate of
the theoretical uncertainties entering the EFT analysis.
\section{Observables}
\label{sec:obs}
\paragraph{Primary observables.}
Although a large number of observables can be constructed from the helicity
amplitudes, the phenomenological strategy of this work is deliberately focused
on a minimal set of three \emph{primary observables}:
the baryonic LFU ratio $\Rlam$, the tau longitudinal polarization $P_\tau$,
and a single angular observable, chosen here as the forward--backward asymmetry
$A_{\rm FB}$.
This set is sufficient to discriminate between vector-like and correlated scalar EFT solutions that remain approximately degenerate in mesonic $b \to c\,\ell\,\bar{\nu}_\ell$ analyses ~\cite{Sakaki2013,Alonso2017,Iguro2024}, while keeping hadronic uncertainties under control.
All additional observables are treated as secondary and are not required to
establish the main physics conclusions.
In this section we define the physical observables used in our analysis.
All observables are expressed in terms of the helicity amplitudes
introduced in Sec.~\ref{sec:ff} and collected explicitly in
Appendix~\ref{app:hel}.
\subsection{Differential decay rate}
\label{subsec:rate}

The differential decay rate for
$\Lambda_b\to\Lambda_c\tau\bar\nu_\tau$ can be written as
\begin{equation}
\frac{d\Gamma}{dq^2}
=
\frac{G_F^2 |V_{cb}|^2}{192\pi^3 m_{\Lambdab}^3}
\, q^2 \sqrt{\lambda(q^2)}\,
\left(1 - \frac{m_\ell^2}{q^2}\right)^2
\mathcal{H}(q^2),
\end{equation}
where $\lambda(q^2)$ is defined in Appendix~\ref{app:hel}, and
$\mathcal{H}(q^2)$ denotes a bilinear combination of helicity amplitudes,
including vector, axial-vector, scalar, and tensor contributions.
The differential rate serves as the common building block for all
subsequent observables and is not used as an independent discriminator
in the EFT analysis.
\subsection{Lepton-flavor universality ratios}
\label{subsec:lfu}
We define the baryonic lepton-flavor universality ratios
\begin{equation}
R_{\Lambdac}
=
\frac{
	\displaystyle
	\int_{m_\tau^2}^{(m_{\Lambdab}-m_{\Lambdac})^2}
	\! dq^2\,
	\frac{d\Gamma(\Lambdab \to \Lambdac \tau \bar\nu_\tau)}{dq^2}
}{
	\displaystyle
	\int_{m_\ell^2}^{(m_{\Lambdab}-m_{\Lambdac})^2}
	\! dq^2\,
	\frac{d\Gamma(\Lambdab \to \Lambdac \ell \bar\nu_\ell)}{dq^2}
},
\end{equation}
with $\ell=e,\mu$.
From a phenomenological perspective, $\Rlam$ plays the role of a
\emph{normalization anchor}.
While $\Rlam$ alone does not fully resolve EFT degeneracies, it provides
a necessary first constraint that restricts the viable parameter space
prior to considering polarization and angular information.
\subsection{Polarization observables}
\label{subsec:pol}
The longitudinal tau polarization is defined as
\begin{equation}
P_\tau(q^2)
=
\frac{
	d\Gamma^{\lambda_\tau=+1/2}/dq^2
	-
	d\Gamma^{\lambda_\tau=-1/2}/dq^2
}{
	d\Gamma^{\lambda_\tau=+1/2}/dq^2
	+
	d\Gamma^{\lambda_\tau=-1/2}/dq^2
}.
\end{equation}
As a result,
$P_\tau(\Lambda_b\to\Lambda_c\tau\bar\nu_\tau)$
provides a decisive discriminator between competing helicity structures
in semitauonic EFT scenarios, including parameter regions that remain
approximately degenerate in integrated rate observables
~\cite{Tanaka2010}.
The sensitivity of $P_\tau$ originates from the helicity structure of
the semileptonic amplitude.
In the massless-lepton limit one recovers
\begin{equation}
P_\tau \to -1,
\end{equation}
reflecting the purely left-handed structure of the Standard Model.
For finite $m_\tau$, scalar and tensor interactions contribute through
helicity-suppressed interference terms proportional to
$m_\tau/\sqrt{q^2}$, thereby modifying the relative weight of positive-
and negative-helicity amplitudes.
\subsection{Angular observables}
\label{subsec:ang}
Angular observables in semileptonic heavy-baryon decays provide direct
access to interference patterns among helicity amplitudes and therefore
probe the chiral structure of the underlying effective interactions
through helicity-interference effects~\cite{Duraisamy2013}.
Among these observables, the lepton forward--backward asymmetry is
particularly sensitive to scalar and tensor contributions that remain
partially degenerate in integrated decay rates.

The forward--backward asymmetry is defined as
\begin{equation}
A_{\rm FB}(q^2)
=
\frac{
	\displaystyle
	\int_0^1 d\!\cos\theta_\ell\,
	\frac{d^2\Gamma}{dq^2 d\!\cos\theta_\ell}
	-
	\int_{-1}^0 d\!\cos\theta_\ell\,
	\frac{d^2\Gamma}{dq^2 d\!\cos\theta_\ell}
}{
	\displaystyle
	\frac{d\Gamma}{dq^2}
},
\end{equation}
where $\theta_\ell$ denotes the angle between the charged lepton and the
$\Lambdac$ baryon in the dilepton rest frame.

The angular distribution can be written in the form
\begin{equation}
\frac{d^2\Gamma}{dq^2\,d\cos\theta_\ell}
=
a(q^2)
+b(q^2)\cos\theta_\ell
+c(q^2)\cos^2\theta_\ell ,
\label{eq:angular_abc}
\end{equation}
which implies
\begin{equation}
\frac{d\Gamma}{dq^2}
=
2a(q^2)+\frac{2}{3}c(q^2),
\end{equation}
and therefore
\begin{equation}
A_{\rm FB}(q^2)
=
\frac{b(q^2)}
{2a(q^2)+\frac{2}{3}c(q^2)} .
\label{eq:AFB_abc}
\end{equation}

The coefficient $b(q^2)$ is odd under
$\cos\theta_\ell\to-\cos\theta_\ell$ and originates from interference
among helicity amplitudes with different Lorentz structure.
Consequently, $A_{\rm FB}$ provides information complementary to both
$\Rlam$ and $P_\tau$, completing a minimal observable basis capable of
separating vector, scalar, and tensor EFT scenarios in
$b\to c\,\tau\bar\nu_\tau$ transitions.
	\section{Numerical analysis and uncertainty propagation}
	\label{sec:num}
	In this section we summarize the numerical inputs used in the analysis,
	define the Standard-Model baseline, specify the priors for the EFT
	Wilson coefficients, and describe the procedure used to propagate
	lattice-QCD form-factor uncertainties to all observables.
	Unless stated otherwise, all Wilson coefficients are taken to be real
	and defined at the scale $\mu = m_b$, and only left-handed neutrinos
	are considered.
	Complex Wilson coefficients could generate additional CP-sensitive
	interference structures in angular observables.
	The present analysis focuses on the CP-conserving limit in order to
	isolate the helicity geometry of the surviving EFT directions.
	Extensions including complex phases and T-odd baryonic observables are
	left for future work.
	\subsection{Mesonic constraints and accepted EFT points}
	\label{subsec:meson_constraints}
	The baryonic predictions shown below are obtained by \emph{propagating}
	EFT points that satisfy the mesonic acceptance criterion of
	Eqs.~\eqref{eq:chi2_meson}--\eqref{eq:chi2_cut_3sigma}.
	Unless stated otherwise we use $\vec O = (R(D),R(D^\ast))$ as the minimal
	mesonic input set, and we treat all Wilson coefficients as real to isolate
	operator fingerprints without introducing CP-violating phases.
	Accepted points are then passed to the heavy-baryon decay framework,
	where we compute the joint predictions for the primary baryonic set
	$\vec X=(R_{\Lambda_c},P_\tau,A_{\rm FB})$ including lattice-QCD form-factor
	covariances as described in Sec.~\ref{subsec:prop}.	
	\subsection{Inputs, priors, and SM baseline}
	\label{subsec:inputs}
	We use the particle masses and lifetimes from the Particle Data Group
	and define the Standard-Model baseline by setting $g_i=0$ in
	Eq.~\eqref{eq:Heff_short}.
	The CKM factor $|V_{cb}|$ is fixed to the exclusive determination.
	Since all observables considered in this work are ratios or normalized
	distributions, their dependence on the precise value of $|V_{cb}|$ is
	numerically negligible.
	Uncertainties on the masses and lifetimes are numerically negligible
	compared to hadronic form-factor uncertainties and are therefore fixed
	to their central values.
	\begin{table}[H]
		\caption{Numerical inputs used in the analysis.}
		\label{tab:inputs}
		\begin{ruledtabular}
			\begin{tabular}{lc}
				Quantity & Value \\
				\hline
				$m_{\Lambdab}$ & $5.6196~\text{GeV}$ \\
				$m_{\Lambdac}$ & $2.2865~\text{GeV}$ \\
				$\tau_{\Lambdab}$ & $1.470~\text{ps}$ \\
				$m_\tau$ & $1.77686~\text{GeV}$ \\
				$\abs{V_{cb}}$ & $0.041$ \\
			\end{tabular}
		\end{ruledtabular}
	\end{table}	
The numerical inputs are taken from the Particle Data Group
Review of Particle Physics~\cite{PDG2024}.
	\subsection{Propagation of lattice uncertainties}
	\label{subsec:prop}
	The lattice-QCD form factors are provided as a set of central values
	and a covariance matrix describing correlations among form factors and
	kinematic points. We treat the form-factor parameters as nuisance
	variables and sample them from a multivariate Gaussian distribution.
	For each draw we compute the full set of helicity amplitudes and all
	derived observables. The quoted theory uncertainties correspond to the
	central $68\%$ interval of the resulting distributions.
	We have verified that the results are stable against increasing the
	number of samples.
\subsection{Helicity structure and EFT implementation}
\label{subsec:consistency}

The sensitivity of baryonic observables to non-Standard-Model chiral
structures originates from the helicity decomposition of the
$b\to c\,\ell\nu$ amplitude. In our conventions, the effective vector
and axial combinations entering the baryonic helicity amplitudes are
\begin{equation}
C_V = 1 + g_{V_L} + g_{V_R},
\qquad
C_A = -1 - g_{V_L} + g_{V_R}.
\label{eq:CVCA}
\end{equation}

The symbolic calculation reproduces the expected helicity structure
of the vector--axial contributions,
\begin{equation}
\Gamma_-(q^2)\propto H_{\rm VA}(q^2),
\qquad
\Gamma_+(q^2)\propto
\frac{m_\tau^2}{2q^2}
\left[
H_{\rm VA}(q^2)+3H_t(q^2)
\right],
\label{eq:GammaHelicityStructure}
\end{equation}
where $H_{\rm VA}$ denotes the sum of transverse and longitudinal
vector--axial helicity amplitudes, while $H_t$ denotes the timelike
helicity contribution.

The corresponding longitudinal tau polarization is
\begin{equation}
P_\tau(q^2)=
\frac{\Gamma_+(q^2)-\Gamma_-(q^2)}
{\Gamma_+(q^2)+\Gamma_-(q^2)}.
\label{eq:PtauStructure}
\end{equation}

In the massless-lepton limit the symbolic implementation gives
\begin{equation}
P_\ell(q^2)\to -1,
\end{equation}
as expected for a purely left-handed charged current.
This provides a stringent validation of the helicity normalization
and sign conventions employed throughout the numerical analysis.

The scalar operators are more transparently expressed through scalar
and pseudoscalar combinations. Using
$P_{L,R}=(1\mp\gamma_5)/2$, one finds
\begin{equation}
g_{S_L}P_L+g_{S_R}P_R
=
\frac12\left(g_S+g_P\gamma_5\right),
\label{eq:ScalarPseudoDecomposition}
\end{equation}
with
\begin{equation}
g_S\equiv g_{S_L}+g_{S_R},
\qquad
g_P\equiv g_{S_R}-g_{S_L}.
\label{eq:gSgP}
\end{equation}

Along the correlated scalar direction used in the numerical scan,
\begin{equation}
g_{S_R}=-g_{S_L}+\epsilon,
\label{eq:CorrelatedScalarDirection}
\end{equation}
one obtains
\begin{equation}
g_S=\epsilon,
\qquad
g_P=\epsilon-2g_{S_L}.
\label{eq:ScalarDirectionResult}
\end{equation}

Thus, for small $\epsilon$, the scalar combination is suppressed
while the pseudoscalar helicity-sensitive contribution remains enhanced.
This analytically explains why correlated scalar solutions can remain
close to vector-like solutions in rate observables while producing
visible shifts in polarization observables such as $P_\tau$.
\subsection{EFT scan regions}
\label{subsec:bench}

Rather than restricting the analysis to isolated benchmark points,
we perform a broad Monte-Carlo exploration of the EFT parameter space.
The scan is organized into four phenomenologically distinct EFT classes:
\begin{itemize}
	\item vector-like solutions $(V)$,
	\item correlated scalar solutions $(S_{LR})$,
	\item tensor solutions $(T)$,
	\item correlated scalar--tensor solutions $(S+T)$.
\end{itemize}
For each class, Wilson coefficients are sampled within physically
motivated ranges around the regions favored by current global analyses
of $b\to c\,\tau\bar\nu_\tau$ anomalies.
All sampled points are subsequently tested against the mesonic
acceptance criterion defined in
Eqs.~\eqref{eq:chi2_meson}--\eqref{eq:chi2_cut_3sigma}.
The vector-like scan explores deviations in $g_{V_L}$ while keeping all
other Wilson coefficients fixed to zero.
The correlated scalar class probes directions satisfying approximately
\begin{equation}
g_{S_R}\simeq -g_{S_L},
\end{equation}
which correspond to approximate flat directions in mesonic observables.
Tensor scenarios are generated by varying $g_T$ within the meson-compatible
region, while the correlated scalar--tensor class simultaneously samples
$(g_{S_L},g_T)$ combinations that remain consistent with current
$R(D)$ and $R(D^\ast)$ measurements.
For every accepted EFT point we compute the baryonic observable set
\begin{equation}
\vec X =
\left(
R_{\Lambda_c},
P_\tau,
A_{\rm FB}
\right),
\end{equation}
including full propagation of lattice-QCD form-factor uncertainties.
This strategy allows the resulting baryonic distributions to represent
continuous EFT manifolds rather than isolated benchmark predictions, similar to the geometric structures observed in global EFT fits
~\cite{Altmannshofer2022}, thereby making the degeneracy structure
directly visible in the observable space.
	\FloatBarrier
	\suppressfloats[t]
	\section{Results}
	\label{sec:results}
	
	We now present the baryonic EFT predictions obtained from the
	meson-compatible parameter regions introduced in
	Sec.~\ref{sec:num}.
	To address this question, we propagate the meson-compatible EFT parameter regions to the baryonic observable space
	\begin{equation}
	\vec X =
	\left(
	R_{\Lambda_c},
	P_\tau,
	A_{\rm FB}
	\right),
	\end{equation}
	using the lattice-QCD helicity form factors discussed in Sec.~\ref{sec:ff}, together with full covariance propagation of hadronic uncertainties.
	The resulting distributions probe not only total decay rates, but also the helicity structure of the underlying charged-current interaction.
	
	Particular emphasis is placed on the polarization observable $P_\tau$, which is highly sensitive to scalar and tensor helicity contributions through interference terms proportional to $m_\tau/\sqrt{q^2}$.
	By contrast, purely vector-like deformations predominantly rescale the Standard-Model amplitude while preserving its left-handed structure.
	\subsection{Sensitivity of baryonic observables}
	\label{subsec:sens_rates}

\onecolumngrid

\begin{figure}[H]
	\centering
	\includegraphics[width=0.95\textwidth]{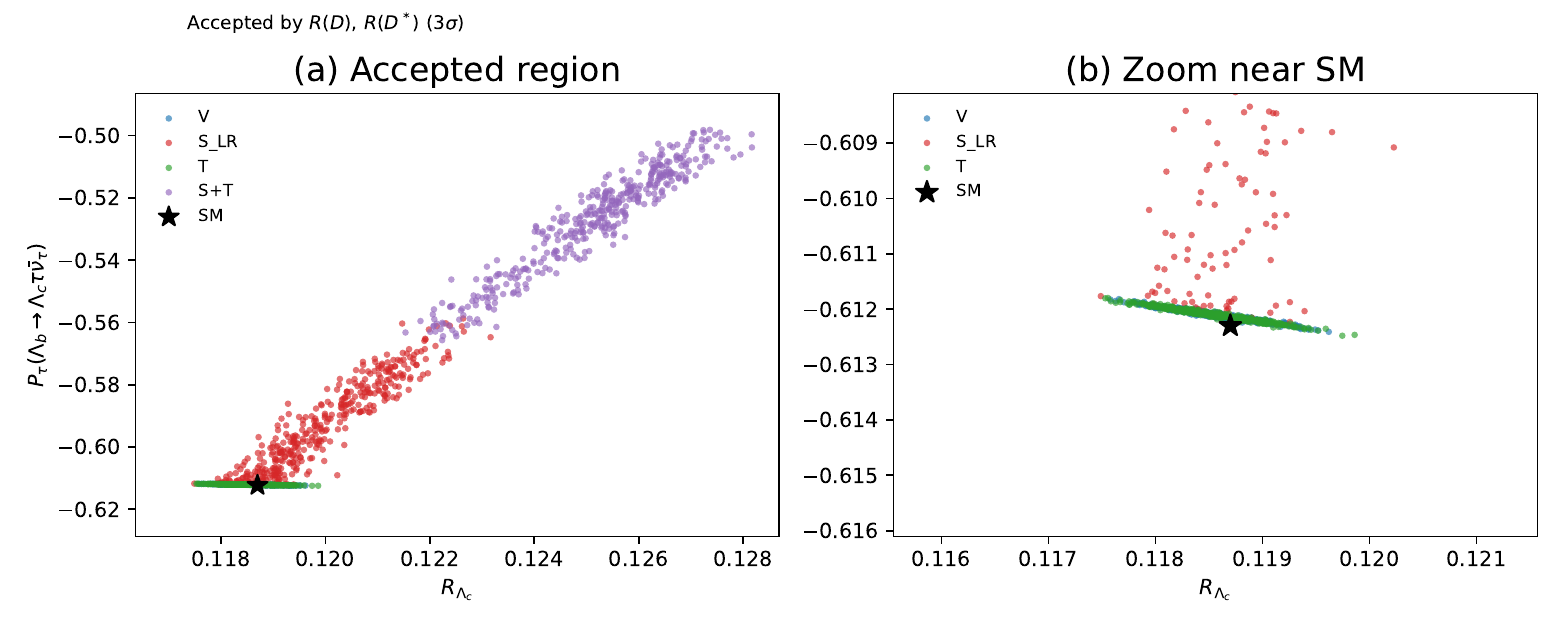}
	\caption{
			Predictions in the $\big(R_{\Lambda_c},\,P_\tau(\Lambda_b\to\Lambda_c\tau\bar{\nu}_\tau)\big)$ plane for EFT parameter points satisfying the mesonic acceptance criterion based on $R(D)$ and $R(D^\ast)$ at the $3\sigma$ level.
			The observables are computed using the helicity-amplitude formalism together with lattice-QCD form factors from Ref.~\cite{Detmold2015}, including full covariance propagation of hadronic uncertainties.
			Panel~(a) shows the full accepted parameter region, while panel~(b) zooms around the Standard-Model point.
			Vector and tensor solutions remain clustered near the Standard-Model prediction, whereas correlated scalar--tensor solutions generate a broad displaced branch associated with helicity-enhanced contributions to the tau polarization.
			The figure demonstrates explicitly how baryonic polarization observables lift EFT degeneracies that remain unresolved in mesonic analyses alone.
	}
	\label{fig:Rlam_Ptau_degeneracy}
\end{figure}

\twocolumngrid
 
   Figure~\ref{fig:Rlam_Ptau_degeneracy} constitutes the central result of this work and demonstrates explicitly how baryonic observables lift the approximate EFT degeneracies that remain unresolved in mesonic analyses of $b\to c\,\tau\bar\nu_\tau$ transitions.
   
   The displayed points correspond to EFT parameter configurations that satisfy the mesonic acceptance criterion defined in Eqs.~\eqref{eq:chi2_meson}--\eqref{eq:chi2_cut_3sigma}, using the experimentally measured values of $R(D)$ and $R(D^\ast)$ at the $3\sigma$ level.
   For each accepted EFT point, the observables
   $\{R_{\Lambda_c},\,P_\tau,\,A_{\rm FB}\}$ are computed using the helicity-amplitude formalism of Sec.~\ref{sec:ff}, together with the lattice-QCD form factors of Ref.~\cite{Detmold2015}.
   Correlations among form factors and kinematic parameters are propagated through the full covariance matrix using multivariate Gaussian sampling, ensuring a consistent treatment of hadronic uncertainties.
   
   Several important features emerge immediately.
   
   First, purely vector-like solutions remain tightly localized around the Standard-Model prediction in the $\big(R_{\Lambda_c},P_\tau\big)$ plane.
   This behavior is expected from the structure of Eq.~\eqref{eq:CVCA}: vector interactions predominantly rescale the Standard-Model charged-current amplitude while preserving its left-handed helicity structure.
   Consequently, both $R_{\Lambda_c}$ and $P_\tau$ remain close to their Standard-Model values even when mesonic observables exhibit sizeable deviations.
   
   Second, tensor solutions display a remarkably similar clustering pattern.
   Although tensor operators modify the helicity decomposition of the decay amplitude, the surviving meson-compatible tensor region remains aligned with the Standard-Model helicity structure in integrated baryonic observables.
   As a result, the $(V,T)$ directions remain approximately degenerate in the projected $\big(R_{\Lambda_c},P_\tau\big)$ plane.
   This surviving degeneracy is itself physically significant, since it identifies a class of EFT deformations that cannot be resolved through integrated LFU ratios alone.
   
   In contrast, correlated scalar--tensor solutions generate a qualitatively distinct structure.
   The broad diagonal branch observed in Fig.~\ref{fig:Rlam_Ptau_degeneracy} originates from the helicity-sensitive pseudoscalar contribution discussed in Eqs.~\eqref{eq:CorrelatedScalarDirection}--\eqref{eq:ScalarDirectionResult}.
   Along this direction the scalar combination remains partially suppressed in rate observables, allowing compatibility with $R(D)$ and $R(D^\ast)$, while the helicity-sensitive interference terms remain enhanced.
   This leads to sizeable shifts in the positive-helicity contribution
   $\Gamma_+(q^2)$ entering Eq.~\eqref{eq:PtauStructure},
   which in turn generates substantial modifications of the integrated
   tau polarization.
  
   To quantify the separation among the surviving EFT classes, we extract
   the mean values of the baryonic observable vector
   \(
   \vec X =
   (R_{\Lambda_c},P_\tau,A_{\rm FB})
   \)
   from the accepted parameter distributions generated in the heavy scan.
   We obtain
   \begin{align}
   \vec X_V
   &=
   (0.11856,\,-0.61210,\,0.23721),
   \\
   \vec X_T
   &=
   (0.11854,\,-0.61210,\,0.23716),
   \\
   \vec X_{S_{LR}}
   &=
   (0.11992,\,-0.59331,\,0.23742),
   \\
   \vec X_{S+T}
   &=
   (0.12513,\,-0.52719,\,0.23832).
   \end{align}
   
   Several important features are immediately visible.
   The vector and tensor scenarios remain nearly indistinguishable at the
   level of integrated baryonic observables, confirming the residual EFT
   degeneracy observed in Fig.~\ref{fig:Rlam_Ptau_degeneracy}.
   By contrast, the correlated scalar--tensor direction produces a sizeable
   shift in the tau polarization while leaving the forward--backward
   asymmetry comparatively stable.
   \subsection{Covariance-aware separation in baryonic observable space}
   \label{subsec:covariance_ellipses}
   
   While the mean vectors provide a compact summary of the surviving EFT
   directions, they do not capture the correlated spread of the accepted
   parameter regions.
   To quantify the geometric separation of the EFT classes in a
   covariance-aware manner, we construct confidence ellipses in the
   $(R_{\Lambda_c},P_\tau)$ observable plane.
   
   For each operator scenario, the mean vector and covariance matrix are
   computed from the accepted EFT points obtained in the heavy scan after
   propagation through the lattice-QCD form-factor covariance matrix.
   The resulting ellipses therefore encode both the intrinsic spread of
   the surviving EFT parameter regions and the correlated baryonic
   response induced by the underlying helicity amplitudes.
   \begin{figure}[H]
   	\centering
   	\includegraphics[width=0.97\columnwidth]
   	{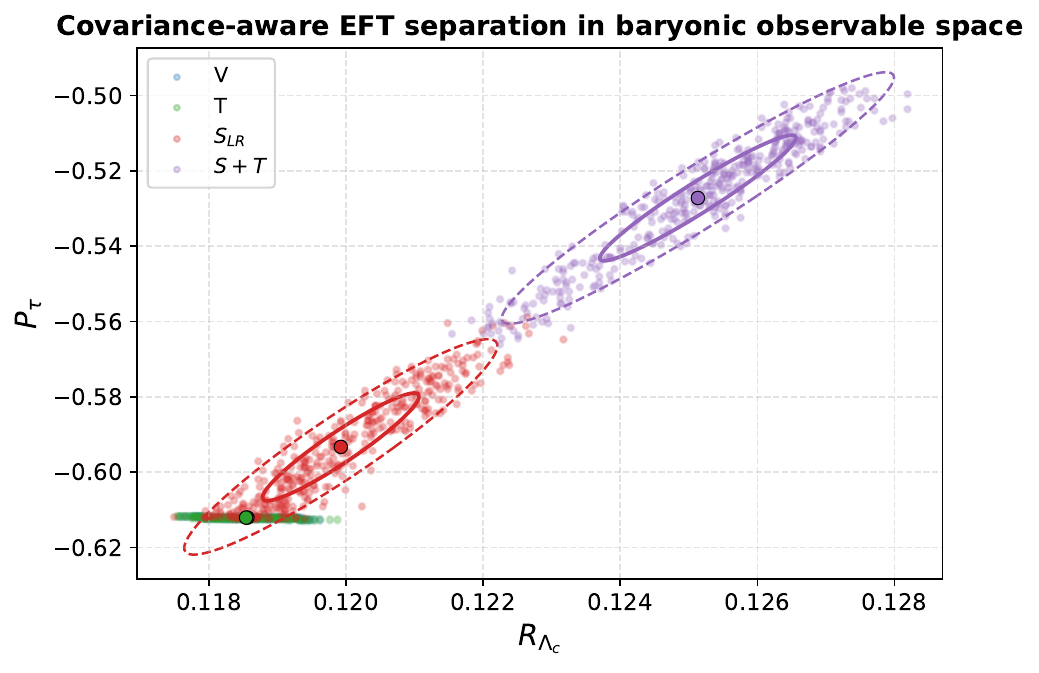}
   	\caption{
   		Covariance-aware separation of meson-compatible EFT solutions in the
   		$(R_{\Lambda_c},P_\tau)$ baryonic observable plane.
   		Points denote accepted EFT solutions from the heavy scan, while solid
   		and dashed ellipses correspond to the associated $1\sigma$ and
   		$2\sigma$ covariance regions.
   		The vector and tensor solutions remain clustered near the
   		Standard-Model-like region, whereas scalar-containing scenarios are
   		displaced along the polarization-sensitive direction.
   		The observed separation demonstrates that baryonic polarization
   		observables provide operator-discriminating information beyond
   		mesonic LFU constraints alone.
   	}
   	\label{fig:CovarianceEllipses}
   \end{figure}
   Figure~\ref{fig:CovarianceEllipses} makes the degeneracy-breaking
   mechanism explicit.
   The vector and tensor hypotheses occupy nearly overlapping covariance
   regions, confirming that integrated rate information alone does not
   fully resolve these EFT directions.
   
   By contrast, scalar-containing scenarios generate a visible
   displacement towards less negative values of $P_\tau$, while the
   correlated scalar--tensor branch exhibits the largest separation from
   the Standard-Model-like region.
   The long axes of the covariance ellipses are predominantly aligned
   with the polarization direction, indicating that the leading EFT
   deformation is controlled by the helicity structure of the tau final
   state rather than by a simple rescaling of the total decay rate.
   
   These results provide a covariance-aware confirmation of the
   qualitative pattern observed in
   Fig.~\ref{fig:Rlam_Ptau_degeneracy} and anticipate the statistical
   separation quantified in the Mahalanobis analysis presented below.
   The substantial overlap of the vector and tensor covariance regions,
   contrasted with the clear displacement of scalar-containing branches,
   demonstrates that polarization-sensitive observables provide the
   dominant baryonic discrimination power among the surviving EFT
   solutions. This behavior is consistent with the small Mahalanobis distance
   obtained below and confirms that the dominant baryonic discrimination
   power originates from scalar-sensitive polarization effects.
   \FloatBarrier
   \subsection{Quantifying EFT separations with covariance-aware distances}
   \label{subsec:mahalanobis}
   To quantify the degree of separation among the surviving EFT solutions
   in a statistically meaningful manner, we compute pairwise Mahalanobis
   distances in the observable space
   \[
   X =
   \left(
   R_{\Lambda_c},
   P_\tau,
   A_{\rm FB}
   \right).
   \]
   The Mahalanobis metric incorporates the expected experimental covariance
   matrix and therefore measures separations relative to the projected
   experimental resolution rather than simple Euclidean distances.
   For two EFT classes $i$ and $j$, the separation is defined as
   \[
   D_{ij}^2
   =
   (X_i-X_j)^T
   \Sigma^{-1}
   (X_i-X_j),
   \]
   where $\Sigma$ denotes the assumed covariance matrix for the baryonic observables. Using the accepted parameter points from the global mesonic scan, we obtain
  \begin{align}
  D(V,T) &\simeq 0.004,
  &
  D(V,S_{LR}) &\simeq 0.63,
  \\
  D(V,S+T) &\simeq 2.79.
  \end{align}
  The hierarchy
  \[
  D(V,T) \ll D(V,S_{LR}) < D(V,S+T)
  \]
  demonstrates that scalar-containing EFT directions are progressively
  separated from the Standard-Model-like region once polarization
  information is incorporated.
  The value $D(V,S+T)\simeq 2.79$ corresponds to a substantial
  separation in the adopted baryonic observable space, indicating that
  scalar--tensor solutions become experimentally distinguishable once
  polarization information is incorporated.
  The results demonstrate quantitatively that pure vector and pure tensor
  solutions remain essentially indistinguishable within the considered
  observable basis, despite surviving mesonic constraints.
  By contrast, correlated scalar--tensor scenarios generate a statistically
  significant displacement driven predominantly by large modifications of
  the tau polarization observable.
   \subsection{Principal discrimination direction}
   \label{subsec:principal_direction} 
   To identify which observable combination carries the dominant
   new-physics information, we perform a principal-direction analysis in
   the baryonic observable space.
   \[
   X =
   \left(
   R_{\Lambda_c},
   P_\tau,
   A_{\rm FB}
   \right).
   \]  
   The covariance matrix constructed from the accepted EFT parameter points
   admits three orthogonal eigen-directions corresponding to the principal
   deformation modes of the surviving parameter space.
   The leading eigenmode accounts for the largest fraction of the variance
   of the surviving EFT manifold and therefore identifies the observable
   direction carrying the strongest discrimination power.
   
   The dominant eigenvector is found to be
   \[
   v_{\rm max}
   \simeq
   -0.08\,\delta R_{\Lambda_c}
   -1.00\,\delta P_\tau
   -0.01\,\delta A_{\rm FB},
   \]
    The corresponding leading eigenvalue accounts for
    \[
    \frac{\lambda_1}
    {\lambda_1+\lambda_2+\lambda_3}
    \simeq 0.9996,
    \]
    indicating that approximately $99.96\%$ of the total variance of the
    accepted EFT manifold is captured by a single principal direction.
    The surviving EFT deformation is therefore effectively one-dimensional
    in the baryonic observable space.
       
    where $\delta X_i$ denotes the deviation of observable $X_i$
    from the corresponding mean value in the accepted EFT ensemble.
    The dominant eigenvector is therefore overwhelmingly aligned with the
    polarization axis, quantitatively establishing
    $P_\tau(\Lambda_b\to\Lambda_c\tau\bar\nu_\tau)$
    as the principal observable driving the separation of the surviving EFT
    directions in baryonic parameter space.
    This result demonstrates that the dominant EFT deformation is governed
    by helicity-sensitive polarization effects rather than by variations of
    the integrated branching ratio or forward--backward asymmetry.
    Consequently, future experimental improvements in the determination of
    $P_\tau$ are expected to provide the largest gains in separating
    otherwise degenerate EFT directions compatible with current mesonic
    constraints.
   \subsection{Differential phase-space deformation of EFT solutions}
   \label{subsec:differential}
\begin{figure}[!b]
   	\centering
   	\includegraphics[width=0.95\columnwidth]{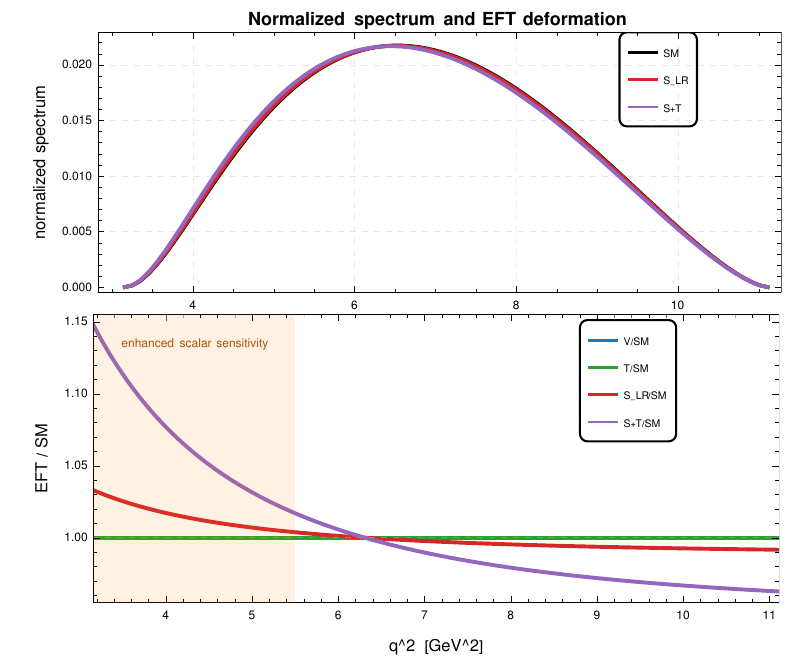}
   	\caption{Differential deformation of the normalized semileptonic spectrum
   		for representative EFT solutions. The upper panel shows the normalized distribution
   		$(1/\Gamma)\, d\Gamma/dq^2$,while the lower panel displays the ratio to the Standard Model prediction. Scalar-containing scenarios generate characteristic low-$q^2$
   		enhancement and high-$q^2$ suppression patterns, whereas the vector- and tensor-like solutions remain nearly spectrally degenerate with the Standard Model prediction.
   		The shaded region highlights the phase-space domain with enhanced
   		sensitivity to scalar contributions.}
   	\label{fig:DiffDeformation}
   \end{figure}

   While integrated baryonic observables already provide substantial
   discrimination power among the surviving EFT solutions, additional information is encoded in the differential structure of the semileptonic decay spectrum, particularly in helicity-sensitive kinematic regions. Differential observables probe the phase-space dependence of the underlying helicity amplitudes and can therefore reveal operator
   structures that remain partially degenerate in integrated decay rates~\cite{Bhattacharya2016,Bordone2020}. In particular, the normalized distribution
   $(1/\Gamma)\,d\Gamma/dq^2$
   retains the phase-space dependence of the underlying helicity amplitudes
   and therefore probes operator-specific interference patterns that may
   remain partially hidden in integrated observables.
   To investigate this effect, we compare representative EFT solutions
   selected from the allowed parameter regions obtained in the global fit.
   The representative EFT curves are obtained by selecting, for each EFT class,
   the accepted point closest to the mean of the corresponding heavy-scan
   distribution in the observable space
   $(R_{\Lambda_c},P_\tau,A_{\rm FB})$, and then evaluating
   $d\Gamma/dq^2$ with the same helicity-amplitude implementation used for
   the integrated observables. Figure~\ref{fig:DiffDeformation}
   shows both the normalized differential spectra and their ratios to the
   Standard Model prediction, following the differential-information strategy
   used in semitauonic EFT studies~\cite{Alonso2017,Datta2017,Bernlochner2018Lambda}.
   We observe that scalar-containing solutions generate characteristic low-$q^2$ enhancement together with moderate suppression at larger $q^2$, reflecting the modified scalar-helicity structure of the decay amplitude. By contrast, the vector- and tensor-like solutions remain nearly spectrally degenerate with the Standard Model prediction over most of phase space, consistent with the approximate null direction identified in the principal-component analysis.
   In contrast, scalar-induced scenarios generate visible low-$q^2$
   deformations, leading to characteristic shape distortions in the
   normalized distribution.
   While the previous differential distributions demonstrate the existence
   of EFT-induced spectral distortions, they do not directly identify where
   the discriminating information is concentrated in phase space.
   To isolate the kinematic origin of the EFT sensitivity, we therefore
   consider the local deformation measure
   \[
   \left|
   \frac{\left(d\Gamma/dq^2\right)_{\rm EFT}}
   {\left(d\Gamma/dq^2\right)_{\rm SM}}
   -1
   \right|,
   \]
   which quantifies the point-by-point departure of the normalized EFT
   spectrum from the Standard Model prediction.
   This representation allows one to determine which phase-space regions
   carry the dominant operator-discrimination power.
   \begin{figure}[!b]
   	\centering
   	\includegraphics[width=0.97\columnwidth]
   	{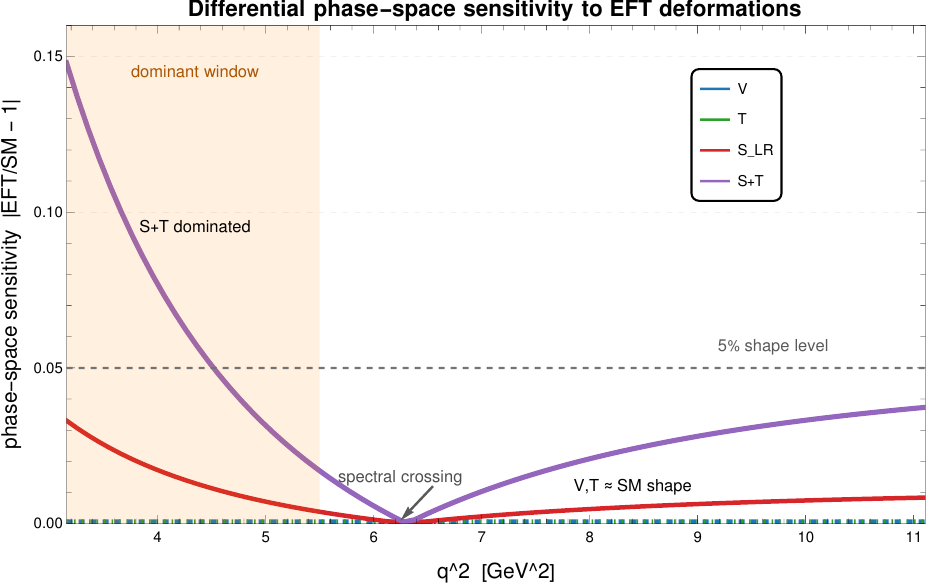}
   	\caption{ Differential phase-space sensitivity to EFT-induced spectral
   		deformations in
   		$\Lambda_b \to \Lambda_c \tau \bar{\nu}_\tau$.
   		Scalar-containing EFT directions generate strongly localized low-$q^2$
   		sensitivity, whereas the surviving vector- and tensor-like solutions
   		remain approximately shape-degenerate with the Standard Model across
   		the full kinematic domain.
   		The spectral-crossing region near
   		$q^2 \simeq 6.2~{\rm GeV}^2$
   		marks the transition between enhancement and suppression regimes of the
   		scalar-induced deformation.
   	}
   	\label{fig:PhaseSpaceSensitivity}
   \end{figure} 
   Figure~\ref{fig:PhaseSpaceSensitivity} demonstrates that the EFT
   discrimination power is highly localized in phase space rather than
   uniformly distributed across the kinematic domain.
   The dominant sensitivity originates from the low-$q^2$ region, where
   scalar helicity amplitudes become comparatively enhanced and induce
   sizable distortions of the normalized spectrum
   ~\cite{Iguro2020}.
   In contrast, the surviving vector- and tensor-like EFT directions remain
   nearly degenerate with the Standard Model shape, indicating that these
   operators predominantly modify the overall normalization while leaving
   the normalized differential spectrum approximately invariant.
   To quantify the cumulative distortion of the normalized semileptonic
   spectrum relative to the Standard Model prediction, we introduce the
   integrated spectral-deformation measure
   \begin{equation}
   \mathcal{I}_{\rm shape}
   =
   \int_{q^2_{\rm min}}^{q^2_{\rm max}}
   dq^2\,
   \left|
   \frac{
   	\left(
   	\dfrac{1}{\Gamma}\dfrac{d\Gamma}{dq^2}
   	\right)_{\rm EFT}
   }{
   	\left(
   	\dfrac{1}{\Gamma}\dfrac{d\Gamma}{dq^2}
   	\right)_{\rm SM}
   }
   -1
   \right|.
   \label{eq:Ishape}
   \end{equation}
   By construction,
   $\mathcal{I}_{\rm shape}=0$
   for a spectrum identical in shape to the Standard Model prediction,
   while larger values indicate increasing operator-induced deformation of
   the normalized phase-space distribution. The observable $\mathcal{I}_{\rm shape}$
   therefore measures the integrated phase-space deformation of the
   normalized differential spectrum induced by a given EFT solution.
 
   Small values correspond to spectra that remain approximately
   shape-degenerate with the Standard Model prediction, whereas larger
   values indicate substantial helicity-dependent spectral distortions.
  
\begin{figure}[H]
	\centering
	\includegraphics[width=0.82\columnwidth]{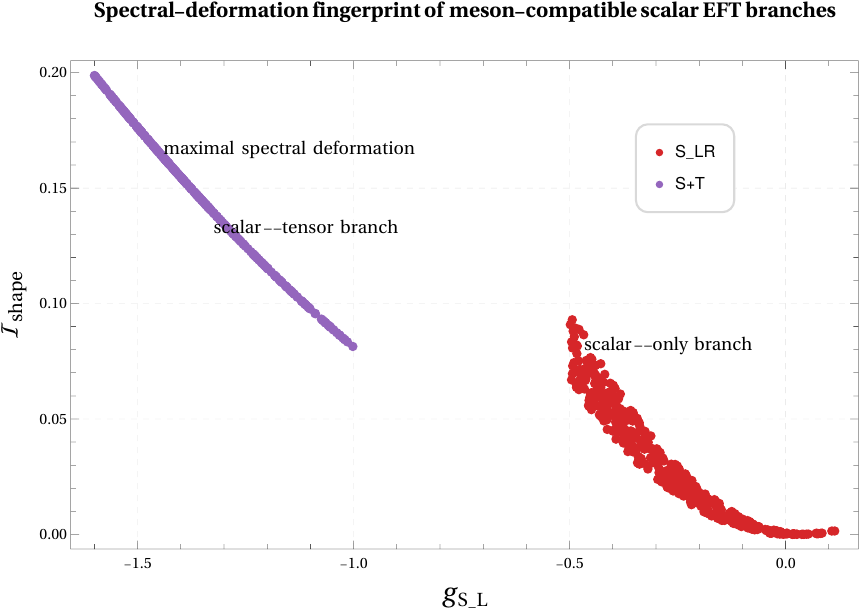}
	\caption{spectral-deformation fingerprint of meson-compatible scalar EFT branches.
		Each point corresponds to an EFT solution compatible with the
		$R(D)$ and $R(D^\ast)$ constraints obtained in the heavy EFT scan.
		The vertical axis shows the integrated spectral-deformation measure
		$\mathcal{I}_{\rm shape}$, defined from the normalized differential
		distribution relative to the Standard Model prediction.
		The surviving scalar-only and correlated scalar--tensor solutions
		populate geometrically separated regions of the EFT manifold,
		demonstrating that differential spectral information provides
		operator-discriminating power beyond integrated LFU observables.
		The scalar--tensor branch exhibits substantially larger spectral
		deformation, particularly in the low-$q^2$ helicity-sensitive region,
		while scalar-only solutions remain closer to the Standard-Model
		spectral geometry.
	}
	\label{fig:ShapeFingerprint}
\end{figure}

Figure~\ref{fig:ShapeFingerprint} provides a complementary geometric view of the surviving scalar EFT directions in the differential observable space. 
Rather than comparing integrated observables alone, the figure probes the cumulative spectral deformation of the normalized semileptonic distribution relative to the Standard Model prediction. The quantity
$\mathcal{I}_{\rm shape}$ therefore acts as a global measure of helicity-induced phase-space
distortion. Several important features emerge immediately. The scalar--tensor solutions populate a broad branch characterized by large spectral deformation, reflecting the enhanced helicity-sensitive interference terms induced by correlated scalar and tensor operators. 

These solutions produce sizable modifications of the low-$q^2$ distribution while remaining compatible with current mesonic constraints. By contrast, scalar-only solutions generate substantially smaller deformations and remain closer to the Standard-Model spectral geometry. This separation demonstrates that the normalized differential spectrum contains complementary operator-discriminating information beyond
integrated LFU ratios. In particular, the figure shows that EFT directions which remain partially degenerate at the level of integrated observables become geometrically separated once differential spectral information is included. The geometric separation of the surviving EFT branches demonstrates
that normalized spectral information resolves operator directions
that remain partially degenerate in integrated observables.
\section{Discussion and interpretation}
\label{sec:disc}
\subsection{Helicity interpretation of EFT degeneracies}
The baryonic observable space
$(R_{\Lambda_c},P_\tau,A_{\rm FB})$
reveals a nontrivial separation of the meson-compatible EFT solutions
identified in the global
$b\to c\,\tau\bar\nu_\tau$
scan.
While vector-like and tensor-like directions remain clustered near the
Standard-Model prediction, scalar-containing solutions generate a
distinct polarization-sensitive branch associated with modified helicity
interference patterns.

The physical origin of this separation is rooted in the helicity
structure of the semileptonic amplitude.
The polarization observable
$P_\tau(\Lambda_b\to\Lambda_c\tau\bar{\nu}_\tau)$
is particularly sensitive to scalar and tensor EFT contributions through
helicity-suppressed interference terms proportional to
$m_\tau/\sqrt{q^2}$.
As a consequence, EFT scenarios that remain approximately degenerate in
mesonic branching ratios become separated once polarization information
is included
~\cite{Sakaki2013,Alonso2017,Celis2017,Iguro2024}.
The accepted EFT distributions reveal two qualitatively distinct
structures.
Vector-like and tensor-like solutions remain localized near the
Standard-Model prediction in the integrated baryonic observable space,
forming an approximately degenerate vector--tensor direction.
This surviving degeneracy is physically significant because it
identifies a class of EFT deformations that cannot be resolved through
integrated LFU observables alone.
By contrast, scalar-containing solutions generate a displaced
polarization-sensitive branch associated with enhanced positive-helicity
contributions to the semileptonic amplitude.

This interpretation is supported consistently by the covariance-aware
Mahalanobis analysis and by the principal-direction decomposition of the
observable manifold.
The dominant deformation eigenvector is found to be aligned almost
entirely with the polarization axis, while the approximate null
direction corresponds to an observable combination that remains only
weakly sensitive to the surviving EFT deformations.
Collectively, these features indicate that
$P_\tau(\Lambda_b\to\Lambda_c\tau\bar{\nu}_\tau)$
provides the dominant baryonic sensitivity to scalar EFT directions.

An additional feature is the relative stability of the forward--backward
asymmetry.
In these regions $A_{\rm FB}$ exhibits comparatively small
variations despite substantial modifications of $P_\tau$.
This behavior reflects the numerical stability of the parity-odd
transverse-helicity interference terms entering the angular coefficient
$b(q^2)$ of Eq.~\eqref{eq:AFB_abc}.
The observable $A_{\rm FB}$ therefore plays a secondary role in the
present EFT geometry, whereas polarization observables dominate the
operator-discrimination power.

Quantitatively, the surviving vector and tensor solutions remain
clustered near
\[
R_{\Lambda_c}\simeq 0.1185,
\qquad
P_\tau\simeq -0.612,
\]
whereas scalar--tensor directions populate a displaced branch extending
toward
\[
R_{\Lambda_c}\simeq 0.125,
\qquad
P_\tau\simeq -0.53.
\]
The resulting hierarchy further emphasizes the enhanced sensitivity of
polarization observables to scalar EFT deformations.
\paragraph{Implications for SMEFT interpretations.}

Although the present analysis is formulated in the low-energy weak
effective theory, the surviving EFT directions can be interpreted as
infrared remnants of SMEFT operators defined above the electroweak
scale.
In particular, correlated scalar and tensor directions may originate
from semileptonic four-fermion operators after electroweak matching and
renormalization-group evolution
~\cite{Jenkins2013,Alonso2014,GonzalezAlonso2017}.
The persistence of the approximately degenerate vector--tensor branch in
the baryonic observable space suggests that certain SMEFT-induced
deformations may remain only weakly constrained by present integrated
LFU measurements.
Future polarization-sensitive baryonic observables could consequently
provide direct information on the ultraviolet helicity structure of
charged-current flavor dynamics.

\subsection{Differential EFT sensitivity}

The differential spectra provide additional sensitivity to the EFT
operator structure beyond that contained in integrated LFU observables.
While vector- and tensor-like solutions remain nearly shape-degenerate
with the Standard Model prediction over most of the kinematic domain,
scalar-containing scenarios generate characteristic distortions of the
normalized semileptonic spectrum.
This behavior reflects a fundamental difference in the underlying
helicity structure of the corresponding EFT operators.

The phase-space sensitivity analysis reveals that the EFT
discrimination power is strongly localized rather than uniformly
distributed across the kinematic domain.
The dominant sensitivity originates from the low-$q^2$ region, where
scalar helicity amplitudes become comparatively enhanced
~\cite{Iguro2020}.
The spectral crossing observed near
\[
q^2 \simeq 6.2~{\rm GeV}^2
\]
marks the transition between enhancement and suppression regimes of
the scalar-induced deformation.

Importantly, the localized spectral distortions cannot be reproduced by
a simple global rescaling of the decay rate~\cite{Becirevic2019}.
Instead, they originate from helicity-dependent interference effects
induced by the EFT operators, providing direct sensitivity to their
Lorentz structure.
Differential baryonic spectra therefore probe aspects of semileptonic
new physics that remain only partially accessible through integrated
observables alone.

An additional perspective is provided by the spectral-deformation
manifold shown in Fig.~\ref{fig:ShapeFingerprint}.
Rather than characterizing EFT solutions solely through integrated
observables, the quantity
$\mathcal{I}_{\rm shape}$
quantifies the cumulative deformation of the normalized semileptonic
spectrum across the full kinematic domain.
The resulting EFT manifold exhibits a clear geometric separation
between scalar-only and correlated scalar--tensor solutions, providing
an independent probe of the helicity structure of the effective
interaction.

Importantly, EFT directions that remain partially degenerate at the
level of integrated LFU observables become separated once normalized
spectral information is incorporated.
The scalar--tensor branch is associated with substantially larger
phase-space deformation, reflecting enhanced helicity-sensitive
interference effects, whereas scalar-only solutions remain closer to
the Standard-Model prediction.
This demonstrates that differential baryonic spectra provide
operator-discriminating information that is largely inaccessible to
purely rate-based analyses and highlights their potential role in
future precision studies of semileptonic flavor dynamics.

\subsection{Experimental prospects}

The principal-direction analysis and projected-precision study indicate
that the future experimental sensitivity will be driven primarily by the
attainable precision on the tau-polarization observable
$P_\tau(\Lambda_b\to\Lambda_c\tau\bar{\nu}_\tau)$.

\onecolumngrid

\begin{figure}[H]
	\centering
	\includegraphics[width=0.92\textwidth]{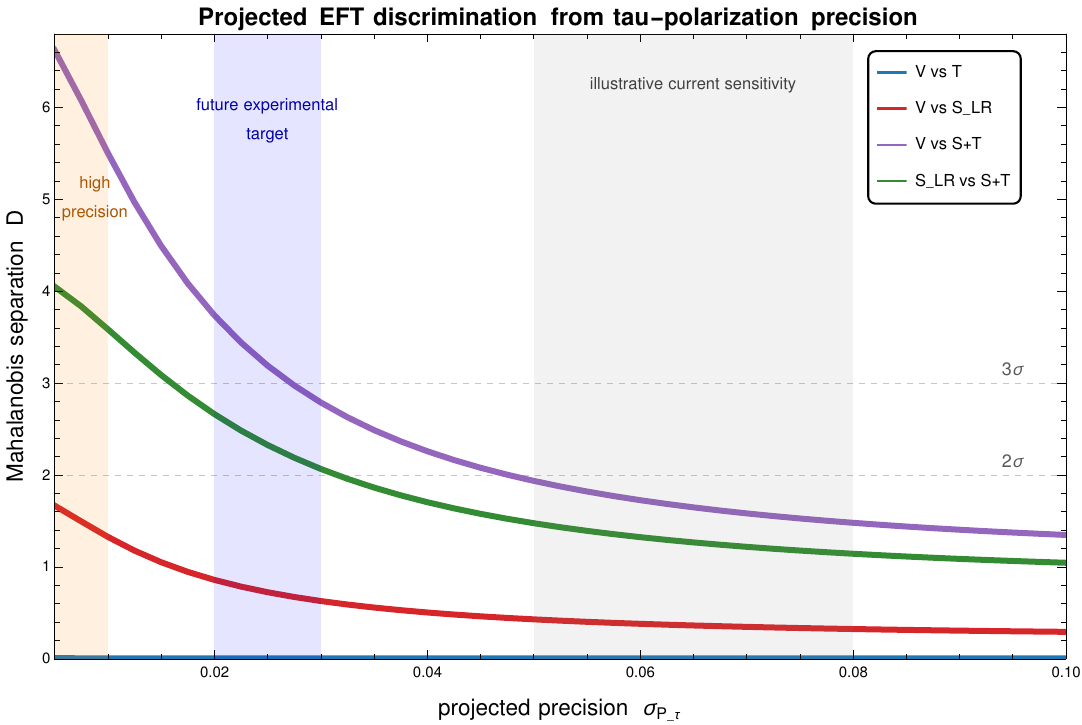}
	\caption{
		Projected EFT discrimination power as a function of the attainable
		precision on
		$P_\tau(\Lambda_b\to\Lambda_c\tau\bar{\nu}_\tau)$.
		The separation is quantified using the Mahalanobis distance in the
		baryonic observable space
		$(R_{\Lambda_c},P_\tau,A_{\rm FB})$.
		The shaded regions indicate representative experimental precision
		benchmarks for future measurements of
		$P_\tau(\Lambda_b\to\Lambda_c\tau\bar{\nu}_\tau)$.
	}
	\label{fig:PtauForecast}
\end{figure}
\FloatBarrier
\twocolumngrid
Figure~\ref{fig:PtauForecast} shows that the EFT discrimination power
depends strongly on the achievable polarization precision.
The vector--tensor direction remains nearly degenerate over the full
precision range considered, indicating that integrated baryonic
observables alone are insufficient to fully resolve this surviving EFT
direction.
Breaking this surviving vector--tensor degeneracy will likely require
observables with enhanced sensitivity to transverse-helicity
interference structures.
Promising candidates include fully differential angular observables,
baryon-spin correlations, and CP-sensitive triple-product asymmetries~\cite{Boer2014},
which probe interference terms absent in integrated LFU ratios.
Such observables could provide access to operator structures that remain
approximately invisible in the present integrated baryonic basis.
By contrast, scalar-containing EFT solutions exhibit rapid growth in
Mahalanobis separation as the polarization uncertainty decreases.
The correlated scalar--tensor branch reaches multi-sigma discrimination
already at moderate precision, demonstrating that future measurements of
$P_\tau(\Lambda_b\to\Lambda_c\tau\bar{\nu}_\tau)$
can efficiently lift EFT degeneracies that remain unresolved in current
mesonic observables
~\cite{LHCbLambdaTau2022,LHCbUpgrade2018,HAMMER2020}.
Experimentally, the relevant baryonic channels are
$\Lambda_b\to\Lambda_c\mu\bar\nu_\mu$
and
$\Lambda_b\to\Lambda_c\tau\bar\nu_\tau$.
The light-lepton differential spectrum has already been measured by
LHCb, while the tauonic channel has been observed with significant
evidence
~\cite{LHCbLambdaMu2017,LHCbLambdaTau2022}.
The LHCb Upgrade and Upgrade II programs therefore provide a realistic
environment for future polarization-sensitive and differential baryonic
measurements.
Our analysis indicates that future experimental efforts should not focus
solely on improving the integrated ratio $R_{\Lambda_c}$.
The strongest EFT sensitivity instead originates from
$P_\tau(\Lambda_b\to\Lambda_c\tau\bar\nu_\tau)$
and from the low-$q^2$ differential spectrum, where scalar sensitivity
is maximal.
Together, these observables transform semileptonic heavy-baryon decays
into a multidimensional probe of the Lorentz and helicity structure of
possible new physics in
$b\to c\tau\bar\nu_\tau$ transitions.
More broadly, the present analysis shows that heavy-baryon decays
elevate the semitauonic flavor-anomaly program from a comparison of
integrated rates to a multidimensional study of helicity geometry,
angular structure, and differential phase-space dynamics.
In this sense, baryonic observables provide a theoretically clean probe
of the operator structure underlying possible lepton-flavor universality
violation in charged-current flavor physics.

	\section{Conclusion}
	\label{sec:concl}
	
	In this work we investigated the extent to which semileptonic
	heavy-baryon decays can resolve the effective-field-theory degeneracies
	that persist in mesonic analyses of
	$b\to c\,\tau\bar\nu_\tau$ transitions.
	Using a dimension-six EFT framework together with lattice-QCD helicity
	form factors and full covariance propagation, we mapped the
	meson-compatible parameter space onto the baryonic observable manifold
	\[
	(R_{\Lambda_c},\,P_\tau,\,A_{\rm FB}),
	\]
	and studied both integrated and differential signatures of the
	surviving EFT directions.
	
	The central result of this analysis is that baryonic observables probe
	the helicity geometry of the semileptonic transition in a way that is
	not accessible through mesonic LFU ratios alone.
	While vector-like and tensor-like EFT solutions remain approximately
	degenerate in integrated observables, scalar-containing directions
	generate characteristic helicity distortions that become directly
	visible through tau polarization and differential spectral information.
	In particular, the observable
	$P_\tau(\Lambda_b\to\Lambda_c\tau\bar\nu_\tau)$
	emerges as the dominant discriminator of scalar and scalar--tensor EFT
	structure.
	
	This conclusion is supported by several independent observations.
	First, the accepted EFT distributions in the
	$(R_{\Lambda_c},P_\tau)$ plane exhibit a pronounced scalar-induced
	polarization branch separated from the Standard-Model-like vector
	region.
	Second, the covariance-aware Mahalanobis analysis shows that
	scalar--tensor solutions become statistically distinguishable once
	polarization information is included, whereas the vector--tensor
	direction remains approximately unresolved.
	Third, the principal-direction analysis demonstrates that the leading
	deformation eigenvector of the accepted EFT manifold is almost entirely
	aligned with the polarization axis, quantitatively establishing
	$P_\tau$ as the primary source of baryonic discrimination power.
	
	The differential analysis provides an additional layer of information.
	We showed that scalar-containing solutions induce localized low-$q^2$
	deformations in the normalized spectrum
	$(1/\Gamma)\,d\Gamma/dq^2$,
	while vector- and tensor-like solutions remain nearly shape-degenerate
	with the Standard Model prediction.
	The dominant EFT sensitivity is concentrated in the low-$q^2$ region,
	with a characteristic spectral crossing near
	$q^2\simeq 6.2~{\rm GeV}^2$,
	indicating that the observed distortions are driven by changes in the
	underlying helicity amplitudes rather than by a simple normalization
	shift of the decay spectrum.
	The spectral-deformation analysis introduced through the observable
	$\mathcal{I}_{\rm shape}$
	further sharpens this conclusion.
	By quantifying the cumulative distortion of the normalized differential
	spectrum relative to the Standard Model prediction, the resulting
	spectral-deformation manifold reveals a clear geometric separation
	between scalar-only and correlated scalar--tensor EFT solutions.
	This demonstrates that differential spectral information carries
	operator-sensitive structure beyond integrated LFU observables and can
	lift EFT degeneracies that remain partially unresolved at the level of
	inclusive rates alone.
	The large deformation observed for the scalar--tensor branch reflects
	enhanced helicity-sensitive interference effects in the low-$q^2$
	region, establishing normalized baryonic spectra as direct probes of
	the underlying helicity geometry of semileptonic new physics.
	
	An important implication of our results is that semileptonic
	heavy-baryon decays provide a powerful probe of the helicity structure
	of charged-current new physics, enabling the resolution of EFT
	degeneracies that survive current mesonic analyses.
	A key advantage of this framework is its theoretical robustness:
	the observables considered here are constructed from lattice-QCD form
	factors with full covariance propagation, allowing the separation of
	competing EFT scenarios to be quantified under controlled theoretical
	uncertainties.
	The complementarity between mesonic and baryonic observables therefore
	extends beyond statistical information.
	While mesonic observables primarily constrain rate-level combinations
	of Wilson coefficients, baryonic polarization and differential spectra
	probe the helicity structure of the underlying interaction, providing
	independent sensitivity to operator directions that remain only weakly
	constrained by integrated LFU measurements.
	
	From an experimental perspective, our analysis identifies a clear
	strategy for future measurements.
	Improving the precision of
	$R_{\Lambda_c}$ alone is not sufficient to fully resolve the surviving
	EFT directions.
	The most powerful probes are instead the tau polarization observable
	and the low-$q^2$ differential spectrum in
	$\Lambda_b\to\Lambda_c\tau\bar\nu_\tau$.
	Measurements of these observables at LHCb Run~3, the LHCb Upgrade, and
	future flavor facilities would provide a direct experimental test of
	whether the present semitauonic anomalies originate from
	helicity-changing scalar interactions or from approximately
	Standard-Model-like vector deformations.
	
	More broadly, our results demonstrate that heavy-baryon decays elevate
	the study of semileptonic flavor anomalies from a program based mainly
	on integrated LFU ratios to a multidimensional analysis of helicity
	structure, angular geometry, and differential phase-space dynamics.
	In this sense, baryonic decays constitute one of the most promising and
	theoretically clean avenues for identifying the operator origin of
	possible lepton-flavor universality violation in charged-current flavor
	physics.
	\begin{acknowledgments}
		The authors sincerely thank Prof.~Driss Khalil and Prof.~Larbi Rahili
		for their continuous support.
		
		We thank W. Detmold, C. Lehner, and S. Meinel for making available the lattice-QCD form-factor results and associated covariance information used in this work through the ancillary material accompanying Ref.~\cite{Detmold2015}.
		
		The numerical computations presented in this work were performed using the computational resources of the MARWAN High Performance Computing Center. We gratefully acknowledge the support and computing facilities provided by the MARWAN infrastructure.

	\end{acknowledgments}

\clearpage
\onecolumngrid
\appendix
\section{Helicity amplitudes}
\label{app:hel}
In this appendix we collect the explicit expressions for the helicity
amplitudes entering the observables discussed in the main text.
The amplitudes are expressed in terms of the helicity form factors
introduced in Sec.~\ref{sec:ff}.
Our conventions follow Refs.~\cite{Bernlochner2018,Datta2019}.
We work in the $\Lambdab$ rest frame and denote by
$\lambda_{\Lambdac}=\pm \tfrac12$ the helicity of the final-state baryon,
and by $\lambda=\pm,0,t$ the polarization of the off-shell $W^\ast$ boson.
The kinematic function
\begin{equation}
\lambda(q^2) =
(m_{\Lambdab}^2 + m_{\Lambdac}^2 - q^2)^2
- 4 m_{\Lambdab}^2 m_{\Lambdac}^2
\end{equation}
is used throughout.
The physical kinematic domain is
$m_\ell^2 \le q^2 \le (m_{\Lambda_b}-m_{\Lambda_c})^2$.
\subsection{Vector and axial-vector helicity amplitudes}
The vector and axial-vector helicity amplitudes are defined in
Eqs.~\eqref{eq:HV_def} and~\eqref{eq:HA_def}.
The helicity amplitudes are obtained by contracting the hadronic
matrix elements with the polarization vectors of the off-shell
$W^\ast$ boson in the dilepton rest frame.
For $\lambda_{\Lambdac}=+\tfrac12$ they read
\begin{align}
H^{V}_{+\frac12,0}
&=
\sqrt{\frac{\lambda(q^2)}{q^2}}\,
\begin{aligned}[t]
\bigg[
& (m_{\Lambdab}+m_{\Lambdac})\, f_+(q^2)
\\
&- \frac{q^2}{m_{\Lambdab}+m_{\Lambdac}}\, f_0(q^2)
\bigg],
\end{aligned}
\\[1mm]
H^{A}_{+\frac12,0}
&=
\sqrt{\frac{\lambda(q^2)}{q^2}}\,
\begin{aligned}[t]
\bigg[
& (m_{\Lambdab}-m_{\Lambdac})\, g_+(q^2)
\\
&+ \frac{q^2}{m_{\Lambdab}-m_{\Lambdac}}\, g_0(q^2)
\bigg],
\end{aligned}
\\[1mm]
H^{V}_{+\frac12,t}
&=
\sqrt{\frac{\lambda(q^2)}{q^2}}\,
(m_{\Lambdab}-m_{\Lambdac})\, f_0(q^2),
\\
H^{A}_{+\frac12,t}
&=
\sqrt{\frac{\lambda(q^2)}{q^2}}\,
(m_{\Lambdab}+m_{\Lambdac})\, g_0(q^2),
\\
H^{V}_{+\frac12,+}
&=
\sqrt{2\,\lambda(q^2)}\,
f_\perp(q^2),
\\
H^{A}_{+\frac12,+}
&=
\sqrt{2\,\lambda(q^2)}\,
g_\perp(q^2).
\end{align}

The amplitudes for $\lambda_{\Lambdac}=-\tfrac12$ follow from parity
relations,
\begin{equation}
H^{V}_{-\lambda_{\Lambdac},-\lambda}
= + H^{V}_{\lambda_{\Lambdac},\lambda},
\qquad
H^{A}_{-\lambda_{\Lambdac},-\lambda}
= - H^{A}_{\lambda_{\Lambdac},\lambda}.
\end{equation}
\subsection{Scalar helicity amplitudes}
The scalar and pseudoscalar helicity amplitudes are not independent
structures, but follow from the contraction of the vector and axial
currents with the momentum transfer
$q^\mu = (p_{\Lambda_b}-p_{\Lambda_c})^\mu$
together with the quark equations of motion,
\begin{equation}
q_\mu\,\bar c \gamma^\mu b
=
(m_b-m_c)\,\bar c\, b,
\qquad
q_\mu\,\bar c \gamma^\mu\gamma_5 b
=
(m_b+m_c)\,\bar c\,\gamma_5 b.
\label{eq:EOMscalar}
\end{equation}
The scalar and pseudoscalar helicity amplitudes can therefore be expressed as
\begin{equation}
H^S_{\lambda_{\Lambda_c}}
=
\frac{q_\mu H^{V,\mu}_{\lambda_{\Lambda_c}}}{m_b-m_c},
\qquad
H^P_{\lambda_{\Lambda_c}}
=
\frac{q_\mu H^{A,\mu}_{\lambda_{\Lambda_c}}}{m_b+m_c}.
\label{eq:HSfromHV}
\end{equation}

Using the equations-of-motion identities above, the scalar helicity
amplitudes reduce to
\begin{align}
H^{S}_{+\frac12}
&=
\frac{\sqrt{\lambda(q^2)}}{m_b - m_c}\, f_0(q^2),
\\
H^{P}_{+\frac12}
&=
\frac{\sqrt{\lambda(q^2)}}{m_b + m_c}\, g_0(q^2),
\end{align}
with analogous relations for $\lambda_{\Lambdac}=-\tfrac12$.
\subsection{Tensor helicity amplitudes}
The tensor helicity amplitudes defined in Eq.~\eqref{eq:HT_def} read.
The tensor indices $(+,-)$ and $(0,t)$ denote transverse and
longitudinal--timelike polarization projections of the antisymmetric
tensor current, respectively.
\begin{align}
H^{T}_{+\frac12,+-}
&=
\sqrt{2\,\lambda(q^2)}\, h_\perp(q^2),
\\
H^{T}_{+\frac12,0t}
&=
\sqrt{\frac{\lambda(q^2)}{q^2}}\,
(m_{\Lambdab}+m_{\Lambdac})\, h_+(q^2).
\end{align}
The remaining tensor amplitudes follow from parity and antisymmetry
relations of the tensor current.
These amplitudes provide direct sensitivity to tensor interactions
beyond the Standard Model.
\subsection{Angular-coefficient decomposition}
\label{app:afbcoeff}
All differential decay rates, polarization observables, and angular
asymmetries discussed in the main text are constructed from bilinear
combinations of the helicity amplitudes listed above.
The EFT sensitivity of the observables therefore originates from
interference among vector, axial, scalar, and tensor helicity
structures with different chiral and polarization properties.

The double-differential decay distribution can be written as
\begin{equation}
\frac{d^2\Gamma}{dq^2\,d\cos\theta_\ell}
=
a(q^2)
+
b(q^2)\cos\theta_\ell
+
c(q^2)\cos^2\theta_\ell .
\label{eq:angularabc}
\end{equation}
The integrated differential rate follows as
\begin{equation}
\frac{d\Gamma}{dq^2}
=
2a(q^2)
+
\frac{2}{3}c(q^2),
\end{equation}
while the forward--backward asymmetry is given by
\begin{equation}
A_{\rm FB}(q^2)
=
\frac{b(q^2)}
{2a(q^2)+\frac{2}{3}c(q^2)}.
\label{eq:afb_final}
\end{equation}
For the effective Hamiltonian of Eq.~\eqref{eq:Heff_short},
the angular coefficients are bilinear combinations of helicity amplitudes.
The coefficient $b(q^2)$ is generated by interference terms among
longitudinal, transverse, and timelike helicity amplitudes and is
therefore particularly sensitive to the chiral structure of the EFT
operators.

The explicit helicity decomposition employed throughout this appendix
makes the polarization and angular sensitivity to different EFT
operators manifest at the amplitude level, providing the basis for the
multidimensional baryonic observables analyzed in the main text.
\end{document}